\documentclass[twocolumn]{revtex4}[12pt]
\usepackage{graphicx}
\begin{document}
\title{Oblique propagation of electrostatic waves in a magnetized EPI plasma}
\author{M. Sarker$^{1,a}$, B. Hosen$^{2}$, M. G. Shah$^{3}$, M. R. Hossen$^{4}$, and A. A. Mamun$^{5}$}
\address{$^{1,2,5}$Department of Physics, Jahangirnagar University, Savar, Dhaka-1342,
Bangladesh\\ $^{3}$Department of Physics, Hajee Mohammad Danesh
Science and Technology University, Bangladesh\\ $^{4}$Department
of General Educational Development, Daffodil
International University, Dhanmondi, Dhaka-1207, Bangladesh\\
Email: $^{a}$sarker.plasma@gmail.com}

\begin{abstract}
A theoretical investigation is carried out to understand the
basic features of nonlinear propagation of heavy ion-acoustic
(HIA) waves subjected to an external magnetic field in an
electron-positron-ion (EPI) plasma which consists of cold
magnetized positively charged heavy ion fluids, and superthermal
distributed electrons and positrons. In the nonlinear regime, the
Korteweg-de Vries (K-dV) and modified K-dV(MK-dV) equations
describing the propagation of HIA waves are derived. The latter
admits a solitary wave solution with both positive and negative
potentials (for K-dV equation) and only positive potential (for
mK-dV equation) in the weak amplitude limit. It is observed that
the effects of external magnetic field (obliqueness), superthermal
electrons and positrons, different plasma species concentration,
heavy ion dynamics and temperature ratio significantly modify the
basic features of solitary waves (SWs). The application of the
results in a magnetized EPI plasma, which occurs in many
astrophysical objects (e.g., pulsars, cluster explosions, and
active galactic nuclei) is briefly discussed.
\end{abstract}
\maketitle
\section{Introduction}
The perusal of electron-positron-ion (EPI) plasmas is very
significant for the understanding of astrophysical as well as
celestial environment. The propagation of linear and nonlinear
waves in such EPI plasmas has drawn a considerable attention to
many authors
\cite{Berezhiani1994,Popel1995,Rizatto1998,Shukla2004,Mushtaq2005,Moslem2007,Tiwari2007,Tribeche2009,Shah2015a,Shah2015b,Shah2015c,Shah2015d}.
In most astrophysical and terrestrial environments (viz. white
dwarfs, Van Allen belts, etc.), positrons coexist with an
admixture of electrons and ions to form EPI plasma, which was
found by many researchers
\cite{Galper1996,Voronov1986,Shapiro1983,Hossen2014s,Hossen2014d,Hossen2017}.
It was noticed that plasma system with positron components behave
differently than regular two component electron-ion (EI) plasmas.
The existence of EPI plasmas has been affirmed in supernovas,
pulsar environments, cluster explosions, active galactic nuclei
(\cite{Begelman1984,Miller1987a,Tribeche2009}), at the center of
Milky Way galaxy \cite{Burns1983}, and in the early universe
\cite{Gibbons1983}.

For modeling purposes, the particle distribution, mostly
envisaged to be  Maxwellian,  when it is very near to equilibrium
in a plasma system. On the basis of isothermal assumption, many
authors studied the ion acoustic waves in EPI plasmas and most of
the studies were concerned with Maxwellian distribution
\cite{Haque2003,Sabry2009,El-Awady2010,Akbari-Moghanjoughi2010,Saeed2010}.
The electrons and positrons which are present in space and
astrophysical plasma environments are not in thermal equilibrium
but highly energetic \cite{Gusev2003,Dwyer2008} due to the effect
of external forces or to wave particle interaction in numerous
space plasma observations \cite{Vocks2003,Gloeckler2006} and
laboratory experiments \cite{Yagi1997,Preische1996}. These
experiments substantiate the existence of accelerated, highly
energetic (superthermal) particles in EPI plasmas. Since the ion
temperature is different from the electron and positron
temperature, so an EPI plasma system which is not in
thermodynamic equilibrium does not follow a Maxwell-Boltzmann
distribution. Plasmas with an excess of superthermal electrons or
positrons elicit a deviation from Maxwellian equilibrium
\cite{Maksimovic1997,Gloeckler2006,Chaston1997}. So
Maxwell-Boltzmann distribution is not appropriate for explicating
the interaction of superthermal particles. However, the plasma
system, containing higher energetic (superthermal) particle with
energies greater than the energies of the particles, exists in
thermal equilibrium that can be fitted more appropriately via the
$\kappa$ (kappa) type of Lorentzian distribution function (DF)
\cite{Vasyliunas1968,Summers1991,Hellberg2009} than via the
thermal Maxwellian DF where the real parameter $\kappa$ measures
the deviation from a Maxwellian distribution (the smaller the
value, the larger the deviation from a Maxwellian, in fact
attained for infinite $\kappa$). Shah \textit{et al.}
\cite{Shah2012} considered an EPI plasma containing inertial
ions, superthermal ($\kappa$ distributed) electrons, and
positrons and studied the effect of positron beam on the ion
acoustic shock waves. Hence we focus on a plasma system with
superthermal particles modelled by a $\kappa$- distribution
\cite{Hellberg2009}.

The three-dimensional (3D) kappa ($\kappa$) velocity distribution
of particles of mass m is of the form
\begin{eqnarray}
&&\hspace*{-5mm}F_k(v)=\frac{\Gamma(k+1)}{(\pi k \omega)^{(3/2)}
\Gamma (k-1/2)}(1+\frac{v^2}{k \omega^2})^{-(k+1)}, \label{1}
\end{eqnarray}

\noindent where, $F_k$ symbolizes the kappa distribution function,
$\Gamma$ is the gamma function, $\omega$ shows the most probable
speed of the energetic particles, given by $\omega =
[(2k-3/k)^{1/2} (k_B T/m)^{1/2} ]$, with $T$ being the
characteristic kinetic temperature and $\omega$ is related to the
thermal speed $V_t = (k_B T/m)^{1/2} $ and, the parameter $k$
represents the spectral index \cite{Cattaert2007} which defines
the strength of the superthermality. The range of this parameter
is $3/2 < k < \infty$ \cite{Alam2013}. In the limit $k
\rightarrow \infty$ \cite{Basu2008,Baluku2012} the kappa
distribution function reduces to the well-known Maxwell-Boltzmann
distribution.

In last few decades, a number of investigations have been made on
the nonlinear propagation of different waves including
ion-acoustic (IA), dust ion-acoustic (DIA), heavy ion-acoustic
(HIA) waves
\cite{Qian1989,Shukla1992,Cairns1996,Shukla2002,Shahmansouri2013a,Shah2015e,Hosen2016}.
Cairns \textit{et al.} \cite{Cairns1996} studied a magnetized
plasma system and observed the effects of external magnetic
field, obliqueness and ion temperature on the amplitude and width
of the IA solitons. Shahmansouri and Alinejad
\cite{Shahmansouri2013a} observed the linear and nonlinear
excitation of arbitrary amplitude IA solitary waves (SWs) in a
magnetized plasma consisting of two-temperature electrons and
cold ions. Jilani \textit{et al.} \cite{Jilani2012} studied the IA
solitons in EPI plasma with non-thermal electrons. Baluku and
Hellberg \cite{Baluku2011} examined the arbitrary amplitude IA SWs
and double layers (DLs) by using the Sagdeev potential approach
in an EPI plasma containing Cairns-distributed (nonthermal)
electrons, Boltzmann positrons and cold ion. Pakzad
\cite{Pakzad2011} investigated a dissipative plasma system with
superthermal electrons and positrons and, observed the effects of
ion kinematic viscosity and the superthermal parameter on the IA
waves. Ghosh \textit{et al.} \cite{Ghosh2013} studied the
nonplanar IA shock waves in a homogeneous unmagnetized EPI plasma
containing superthermal electrons, positrons, and singly charged
hot positive ions. The existence of positively and negatively
charged heavy ions, dust of opposite polarities has been shown to
exist in astrophysical environments
\cite{Ellis1991,Chow1993,Shukla2002}. Baluku \cite{Baluku2010}
studied the behaviour and existence of DIA waves in a plasma
having both polarities of dust. The differences between HIA waves
and DIA waves in plasmas are as follows: (a) the frequency of HIA
waves is much smaller than that of DIA waves; (b) in DIA waves,
static dust grains participate only in maintaining the equilibrium
charge neutrality condition, whereas in HIA waves, mobile heavy
ions provide the necessary inertia; and (c) in DIA waves, the
inertia is provided by the light ions mass, whereas in HIA waves,
inertia comes from the heavy ion mass. Hossen \textit{et al.}
\cite{Hossen2014} examined the characteristics of HIA solitary
structures associated with the nonlinear electrostatic
perturbations in an unmagnetized, collisionless dense plasma.

To the best of our knowledge, no theoretical investigations have
been made on the propagation of HIA waves by deriving the
magnetized Korteweg-de Vries (K-dV) and magnetized modified K-dV
(mK-dV) equations to understand the nonlinear exitatations in
plasmas containing superthermal electrons and positrons, and heavy
ions. In addition, the solitary wave solution of magnetized K-dV
and mK-dV equations as well as other parameters like polarity,
nonlinearity and dispersion coefficients with amplitude and width
of the solitary structures have been studied for different
relevant parameters.
The manuscript is organized as follows. The
basic equations are provided in section \ref{Sec:2}. Two
different types of nonlinear equations, namely K-dV and mK-dV are
derived and analyzed analytically and numerically in section
\ref{Sec:3}. A brief discussion is finally presented in section
\ref{Sec:4}.
\section{Basic Equations}
\label{Sec:2} We consider a three component magnetized plasma
system containing positively charged heavy ions and kappa
distributed electrons and positrons with two distinct
temperatures $T_e$ and $T_p$ . Therefore, at equilibrium
condition, $Z_hn_{h0}=n_{e0}+n_{p0}$, where $Z_h$ ($n_{h0}$)  is
charge state (equilibrium number density) of the heavy ion
species, and $n_{e0}$ ($n_{p0}$) is the equilibrium electron
(positron) number density at temperature $T_e$ ($T_p$). The
dynamics of the heavy ion-acoustic waves, whose phase speed is
much smaller (larger) than electron (heavy ion) thermal speed, is
described by the normalized equations in the form
\begin{eqnarray}
&&\frac{\partial n_h}{\partial t} +
\nabla.({n_h}{\textbf{u}_h})=0,
\label{2}\\
&&\frac{\partial \textbf{u}_h}{\partial t} +
({\textbf{u}_h}.\nabla)u_h=- \nabla \phi+
\alpha({\textbf{u}_h}\times{\hat{z}}),
\label{3}\\
&&\nabla^2\phi=\mu_0n_e+\mu_1n_p-n_h, \label{4}
\end{eqnarray}
where $n_h$ is the heavy ion number density normalized by
$n_{h0}$; $\textbf u_h$ is the heavy ion fluid speed normalized
by $C_h=(Z_hk_BT_e/m_h)^{1/2}$ (with $k_B$ being the Boltzmann
constant, and $m_h$ being the heavy ion mass);  $\phi$ is the
electrostatic wave potential normalized by $k_BT_e/e$ (with $e$
being the magnitude of the charge of an electron); $\alpha
=\omega_{hc}/\omega_{ph}$ (with $\omega_{ch}=Z_heB_0/m_hc$ being
the heavy ion-cyclotron frequency, $B_0$ being the magnitude of
the external static magnetic field, $c$ is the speed of light in
vacuum, and ${\omega_{ph}}=(4 \pi n_{h0}Z_h^2e^2/m_h)^{1/2}$ being
the heavy ion plasma frequency); $\mu_0=\mu/(1+\mu)$,
$\mu_1=1/(1+\mu)$, and $\mu=n_{e0}/n_{p0}$; $n_e$ ($n_p$) is the
number density of electron (positron); time variable is
normalized by $\omega_{ph}^{-1}$, and the space variable is
normalized by $\lambda_{Dm}= C_h/\omega_{ph}$. We note that the
external magnetic field ${\bf B_0}$ is acting along the
z-direction (i.e. ${\bf B_0}=\hat{z}B_0$, where $\hat{z}$ is the
unit vector along the z-direction). The normalized electron and
positron number densities $n_e$ and $n_p$ are, respectively given
by
\begin{eqnarray}
&&n_e=\left(1-\frac{\phi}{\kappa_{1}-\frac{3}{2}}\right)^{-\kappa_{1}
+\frac{1}{2}},
\label{5}\\
&&n_p=\left(1-\frac{\sigma\phi}{\kappa_{2}-\frac{3}{2}}\right)^{-\kappa_{2}+\frac{1}{2}},
\label{6}
\end{eqnarray}
where $\sigma=T_e/T_p$, $\kappa_1$ ($\kappa_2$) is the spectral
index of superthermal electron (positron).
\section{Nonlinear Equations}
\label{Sec:3} To study nonlinear propagation, we now consider
different orders of nonlinearity by deriving and analyzing K-dV
and MK-dV equations to identify the basic features of HIA SWs
formed a magnetized space plasma system containing dynamical
heavy ions, and kappa distributed electrons and positrons of two
distinct temperatures.
\subsection{K-dV Equation}
To derive the K-dV equation, we use the reductive perturbation
method which leads to the stretched co-ordinates \cite{Mamun1998}:
\begin{eqnarray}
&&\xi=\epsilon^{1/2}(l_xx+l_y y+l_z z -V_pt),
\label{7}\\
&&\tau={\epsilon}^{3/2}t, \label{8}
\end{eqnarray}
where $V_p$ is the phase speed of the HIA SWs, $\epsilon$ is a
smallness parameter measuring the weakness of the dispersion
$(0<\epsilon<1)$, and $l_x$, $l_y$, and $l_z$ are the directional
cosines of the wave vector ${\bf k}$ (so that
$l_x^2+l_y^2+l_z^2=1$), as well as leads to the expansion of the
perturbed quantities $n_h$, $u_h$, and $\phi$ in power series of
$\epsilon$:

\begin{eqnarray}
&&n_h=1+\epsilon n_h^{(1)}+\epsilon^{2}n_h^{(2)}+ \cdot \cdot
\cdot, \label{9}\\
&&u_{hx,y}=0+\epsilon^{3/2}
u_{hx,y}^{(1)}+\epsilon^{2}u_{hx,y}^{(2)}+\cdot \cdot \cdot,
\label{10}\\
&&u_{hz}=0+\epsilon u_{hz}^{(1)}+\epsilon^{2}u_{hz}^{(2)}+\cdot
\cdot \cdot,
\label{11}\\
&&\phi=0+\epsilon\phi^{(1)}+\epsilon^{2}\phi^{(2)}+\cdot \cdot
\cdot    . \label{12}
\end{eqnarray}
Now, substituting Eqs. (\ref{7})$-$(\ref{12}) into Eqs.
(\ref{2})$-$ (\ref{4}), and then taking the terms containing
$\epsilon^{3/2}$ from Eqs. (\ref{2}) and (\ref{3}), and $\epsilon$
from Eq. (\ref{4}), we obtain
\begin{eqnarray}
&&u_{hx}^{(1)}=-\frac{l_y}{\alpha}
\frac{\partial\phi^{(1)}}{\partial\xi},
\label{13}\\
&&u_{hy}^{(1)}=\frac{l_x}{\alpha}
\frac{\partial\phi^{(1)}}{\partial\xi},
\label{14}\\
&&u_{hz}^{(1)}=l_z \frac{\phi^{(1)}}{V_p},
\label{15}\\
&&n_{h}^{(1)}=l_z^2\frac{\phi^{(1)}}{V_p^2},
\label{16}\\
&&V_p=\frac{l_z}{\sqrt{\mu_0c_1-\mu_{1}d_1}}, \label{17}
\end{eqnarray}
where
\begin{eqnarray}
&&c_1=\frac{(2\kappa_{1}-1)}{2\kappa_{1} -3},
\label{18}\\
&&d_1=\frac{(2\kappa_{2} -1)\sigma}{2\kappa_{2} -3}. \label{19}
\end{eqnarray}

We note that Eq. (\ref{17}) describes the linear dispersion
relation for the propagation of the HIA SWs  in the magnetized
plasma under consideration and that $l_z=\cos\delta$ (where
$\delta$ is the angle between the directions of external magnetic
field and wave propagation). To the next higher order of
$\epsilon$, we again substitute Eqs. (\ref{7})$-$(\ref{12}) into
Eqs. (\ref{2}), z-component of Eq. (\ref{3}), and Eq.(\ref{4}) and
take the terms containing $\epsilon^{5/2}$ from Eq. (\ref{2}) and
z-component of Eq. (\ref{3}), and $\epsilon^2$ from Eq. (\ref{4}).
We then use Eqs. (\ref{13})$-$(\ref{17}) to obtain a set of
equations in the form
\begin{eqnarray}
&&\frac{\partial n_h^{(1)}}{\partial \tau}-v_p \frac{\partial
n_h^{(2)}}{\partial \xi}+l_x\frac{\partial u_{hx}^{(2)}}{\partial
\xi}\nonumber\\&&+l_y\frac{\partial u_{hy}^{(2)}}{\partial
\xi}+l_z\frac{\partial u_{hz}^{(2)}}{\partial
\xi}+l_z\frac{\partial {(n_h^{(1)}u_{hz}^{(1)})}}{\partial \xi}=0,
\label{20}\\
&&\frac{\partial u_{hz}^{(1)}}{\partial \tau}-V_p\frac{\partial
u_{hz}^{(2)}}{\partial \xi}+l_z u_{hz}^{(1)}\frac{\partial
u_{hz}^{(1)}}{\partial \xi}+l_z\frac{\partial
{\phi_2}}{\partial\xi}=0,
\label{21}\\
&&\frac{\partial^2 \phi_1}{\partial
\xi^2}=\mu_{0}c_1\phi^{(2)}+\mu_{0}c_2{\phi^{(1)}}^2\nonumber\\&&-\mu_{1}d_1\phi^{(2)}-\mu_{1}d_2{\phi^{(1)}}^2-n_h^{(2)}.
\label{22}\
\end{eqnarray}
where
\begin{eqnarray}
&&c_2=\frac{(2\kappa_{1}-1)(2\kappa_{1}+1)}{2(k_{1}-3)^2},
\label{23}\\
&&d_2=\frac{(2\kappa_{2}-1)(2\kappa_{2}+1)\sigma^2}{2(\kappa_{2}-3)^2}.
\label{24}
\end{eqnarray}

 On the other-hand, substituting  Eqs. (\ref{7})$-$(\ref{12})
into $x-$ and $y-$ components of Eq. (\ref{3}), and taking the
terms containing $\epsilon^2$, we get
\begin{eqnarray}
&&u_{hy}^{(2)}=\frac{l_yV_P}{\alpha^2}\frac{\partial^2\phi^{(1)}}{\partial\xi^2},
\label{25}\\
&&u_{hx}^{(2)}=\frac{l_xV_P}{\alpha^2}\frac{\partial^2\phi^{(1)}}{\partial\xi^2}.
\label{26}
\end{eqnarray}
Now combining the equation (\ref{20})$-$(\ref{26}), we have a
equation of the form

\begin{eqnarray}
&&\frac{\partial\phi^{(1)}}{\partial \tau} + A_1 \phi^{(1)}
\frac{\partial \phi^{(1)}}{\partial \xi}+ B_1 \frac{\partial^3
\phi^{(1)}}{\partial \xi^3}=0. \label{27}
\end{eqnarray}
where
\begin{eqnarray}
&&A_1=\frac{3l_z^2}{2V_p}-V_p,
\label{28}\\
&&B_1=\frac{V_p^3}{2l_z^2}\left[1+\frac{(1-l_z^2)}{\alpha^2}\right].
\label{29}\
\end{eqnarray}
Equation (\ref{27}) is the K-dV equation describing the nonlinear
dynamics of the HIA SWs. Now, using the appropriate boundary
conditions, viz. $\phi=0$, $d\phi/d\xi=0$, and $d^2\phi/d\xi^2=0$
at $\xi\rightarrow \pm \infty$,  the stationary solitary wave
solution of Eq. (\ref{27}) is given by
\begin{eqnarray}
{\rm \phi^{(1)}}=\rm \phi_m{\rm[sech^{2}}(\frac{\xi}{\Delta})],
\label{solK-dV}
\end{eqnarray}
where $\phi_m=3u_{0}/A_1$ is the amplitude, and
$\Delta=(4B_1/u_{0})^{1/2}$ is the width of the HIA SWs.
\begin{figure}[t!]
\centerline{\includegraphics[width=8cm]{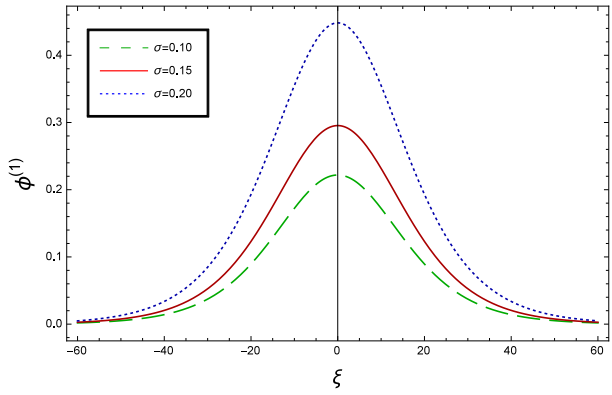}} \caption{(Color
online) The electrostatic solitary potential profiles (ESPPs)
with $\phi^{(1)}>0$ for $\mu>\mu_{c}$, $u_0=0.01$, $\mu=2.98$,
$\kappa_1=20$, $\kappa_2=3$, $\delta=15$, $\alpha =0.5$,
$\sigma=0.10$ (dashed curve), $\sigma=0.15$ (solid curve), and
$\sigma=0.20$ (dotted curve).}  \label{Fig1}
\end{figure}
\begin{figure}[t!]
\centerline{\includegraphics[width=8cm]{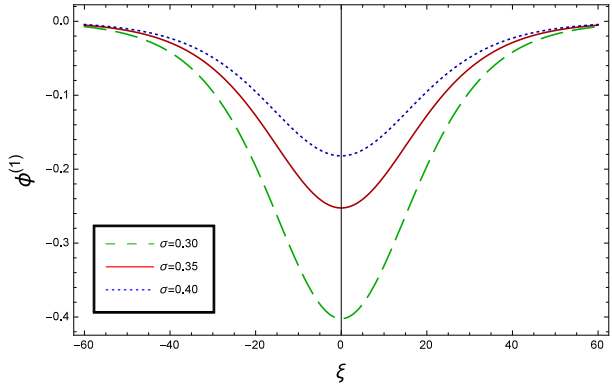}} \caption{(Color
online) The ESPPs with $\phi^{(1)}<0$ for $\mu<\mu_{c}$,
$u_0=0.01$, $\mu=2.63$, $\kappa_1=20$, $\kappa_2=3$, $\delta=15$,
$\alpha =0.5$, $\sigma=0.30$ (dashed curve), $\sigma=0.35$ (solid
curve), and $\sigma=0.40$ (dotted curve).} \label{Fig2}
\end{figure}
\begin{figure}[t!]
\centerline{\includegraphics[width=8cm]{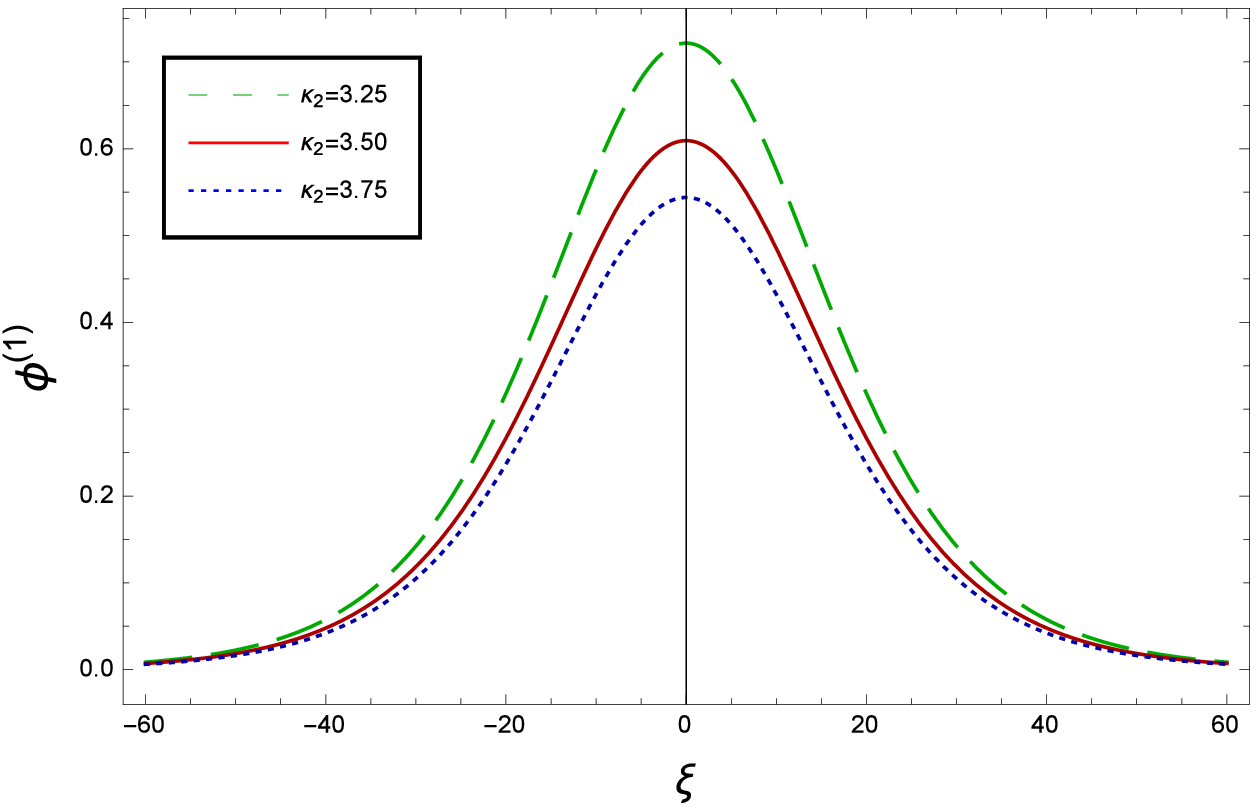}} \caption{(Color
online) The ESPPs with $\phi^{(1)}>0$ for $\mu>\mu_{c}$,
$u_0=0.01$, $\sigma = 0.25$, $\mu=2.98$, $\kappa_1=20$,
$\delta=15$, $\alpha =0.5$, $\kappa_{2}=3.25$ (dashed curve),
$\kappa_{2}=3.50$ (solid curve), and $\kappa_{2}=3.75$ (dotted
curve).} \label{Fig3}
\end{figure}
\begin{figure}[t!]
\centerline{\includegraphics[width=8cm]{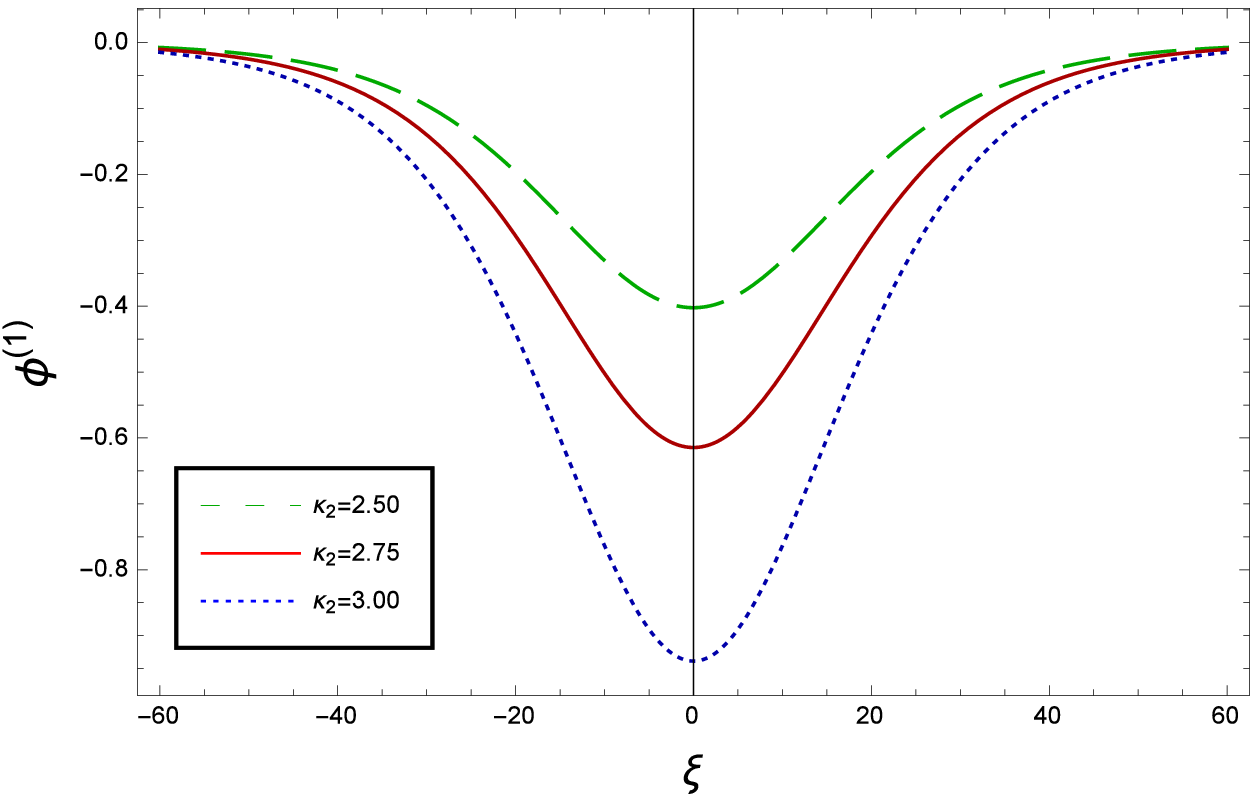}} \caption{(Color
online) The ESPPs with $\phi^{(1)}<0$ for $\mu<\mu_{c}$,
$u_0=0.01$, $\sigma = 0.25$, $\mu=2.63$, $\kappa_1=20$,
$\delta=15$, $\alpha =0.5$, $\kappa_{2}=2.50$ (dashed curve),
$\kappa_{2}=2.75$ (solid curve), and $\kappa_{2}=3.00$ (dotted
curve).} \label{Fig4}
\end{figure}
\begin{figure}[t!]
\centerline{\includegraphics[width=8cm]{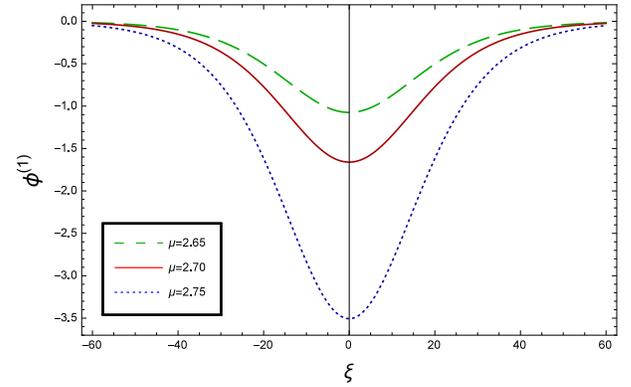}} \caption{(Color
online) The ESPPs with $\phi^{(1)}<0$ for $\mu<\mu_{c}$,
$u_0=0.01$, $\sigma = 0.25$, $\kappa_1=20$, $\kappa_{2}=3$,
$\delta=15$, $\alpha =0.5$, $\mu=2.65$ (dashed curve), $\mu=2.70$
(solid curve), and $\mu=2.75$ (dotted curve).} \label{Fig5}
\end{figure}
\begin{figure}[t!]
\centerline{\includegraphics[width=8cm]{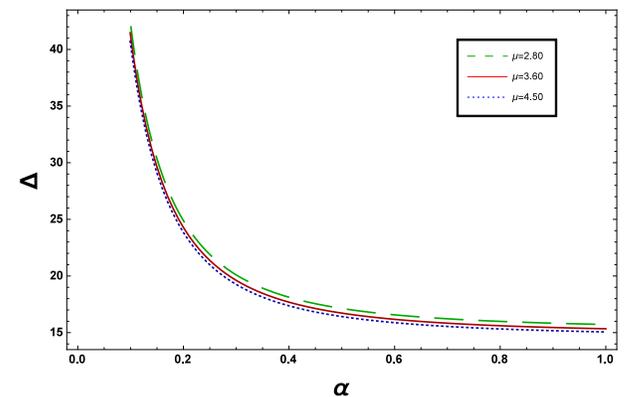}} \caption{(Color
online) The ESPPs with $\phi^{(1)}>0$ for $\mu>\mu_{c}$,
$u_0=0.01$, $\delta = 15$, $\sigma=0.25$, $\kappa_1=20$,
$\kappa_{2}=3$, $\mu=2.80$ (dashed curve), $\mu=3.6$ (solid
curve), and $\mu=4.50$ (dotted curve).} \label{Fig6}
\end{figure}
\begin{figure}[t!]
\centerline{\includegraphics[width=8cm]{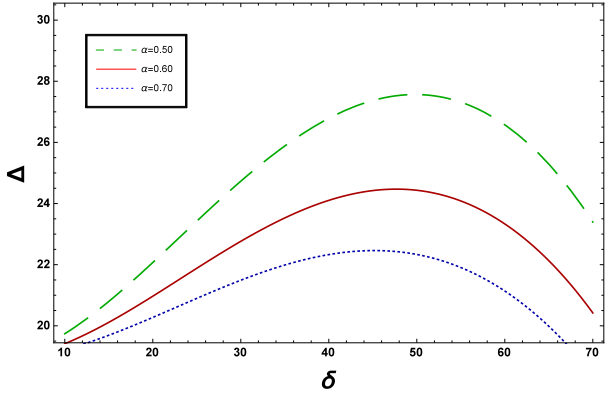}} \caption{(Color
online) The width of the ESPPs for $\mu>\mu_{c}$, $u_0=0.01$,
$\sigma = 0.25$, $\mu=2.98$, $\kappa_1=20$, $\kappa_{2}=3$,
$\alpha=0.5$ (dashed curve), $\alpha=0.6$ (solid curve), and
$\alpha=0.7$ (dotted curve).} \label{Fig7}
\end{figure}
To obtain the basic features (viz. polarity, amplitude, and width)
of the ESPPs, we have numerically analyzed the solution, Eq.
(\ref{solK-dV}) for different plasma situations. The results are
displayed in Figs. \ref{Fig1}$-$\ref{Fig7}, which clearly indicate
that (i) the ESPPs with $\phi^{(1)}>0$ ($\phi^{(1)}<0$) exist for
$\mu>\mu_{c}$ ($\mu<\mu_{c}$) as shown in Figs.
\ref{Fig1}$-$\ref{Fig5}; (ii) the amplitude and width of the ESPPs
[with both $\phi^{(1)}>0$ and $\phi^{(1)}<0$] increase (decrease)
with the increase in $\sigma$ ($\kappa_{2}$) as shown in Figs.
\ref{Fig1}-\ref{Fig4}; (iii) the amplitude of the ESPPs [with
$\phi^{(1)}<0$ for $\mu<\mu_{c}$] decreases with the increase in
$\mu$ as shown in Fig. \ref{Fig5}; (iv) the width of the ESPPs
[with both $\phi^{(1)}>0$] also decreases slightly with the
gradually increase in $\mu$ and $\alpha$ as shown in Fig.
\ref{Fig6}; (v) the width of the ESPPs [with $\phi^{(1)}>0$]
increases (decreases) with the increase in $\delta$ for its lower
(upper) range, but it decreases with the increase in $\alpha$ as
shown in Fig. \ref{Fig7}.

\subsection{MK-dV Equation}
To derive the MK-dV equation we use the same stretched
co-ordinates defined by  Eqs. (\ref{7}) and (\ref{8}), but the
different types of expansion of the dependent variables:
\begin{eqnarray}
&&n_h=1+\epsilon^{1/2} n_h^{(1)}+\epsilon n_h^{(2)}+\epsilon^{3/2}
n_h^{(3)} +\cdot \cdot
\cdot, \label{30}\\
&&u_{hx,y}=0+\epsilon
u_{hx,y}^{(1)}+\epsilon^{3/2}u_{hx,y}^{(2)}+\epsilon^{2}u_{hx,y}^{(3)}+\cdot
\cdot \cdot,
\label{31}\\
&&u_{hz}=0+\epsilon^{1/2} u_{hz}^{(1)}+\epsilon
u_{hz}^{(2)}+\epsilon^{3/2} u_{hz}^{(3)}+\cdot \cdot \cdot,
\label{32}\\
&&\phi=0+\epsilon^{1/2}\phi^{(1)}+\epsilon\phi^{(2)}+\epsilon^{3/2}\phi^{(3)}\cdot
\cdot \cdot, \label{33}
\end{eqnarray}
Now, substituting Eqs. (\ref{7}), (\ref{8}) and
(\ref{30})$-$(\ref{33}) into Eqs. (\ref{2})$-$(\ref{4}), and then
taking the terms containing $\epsilon$ from Eq. (\ref{2}) and
z-component of Eq. (\ref{3}), and $\epsilon^{1/2}$ from Eq.
(\ref{4}), we find the expressions for $n_{h}^{(1)}$,
$u_{hx}^{(1)}$, $u_{hy}^{(1)}$, $u_{hz}^{(1)}$, $V_p$,
$u_{hx}^{(2)}$, and $u_{hy}^{(2)}$ which have already been given
by Eqs. (\ref{13})$-$(\ref{17}), (\ref{25}), and (\ref{26}). To
the next higher order of $\epsilon$, again we substitute Eqs.
(\ref{7}), (\ref{8}) and (\ref{30})$-$(\ref{33}) into Eqs.
(\ref{2})$-$(\ref{4}), and take the terms containing
$\epsilon^{3/2}$ from  Eq. (\ref{2}) and  the z-component of Eq.
(\ref{3}), and $\epsilon$ from Eq. (\ref{4}). Then using the
expressions for $n_{h}^{(1)}$, $u_{hx}^{(1)}$, $u_{hy}^{(1)}$,
$u_{hz}^{(1)}$, $V_p$, $u_{hx}^{(2)}$, and $u_{hy}^{(2)}$ in
these higher order equations, we obtain a set of equations:
\begin{eqnarray}
&&\hspace*{-38mm}u_{hz}^{(2)}=\frac{l_z^3{\phi^{(1)}}^2}{2{V_{p}}^3}+\frac{l_z
{\phi^{(2)}} }{V_p},
\label{34}\\
&&\hspace*{-38mm}n_h^{(2)}=\frac{3l_z^4
{\phi^{(1)}}^2}{2V_p^4}+\frac{l_z^2 {\phi^{(2)}}}{V_p^2},
\label{35}\\
&&\hspace*{-38mm}\rho^{(2)}=-\frac{1}{2}A{\phi^{(1)}}^2=0,
\label{36}\
\end{eqnarray}
where,
$$A=\left[\frac{3l_z^4}{2V_p^4}-(\mu_{0}c_2-\mu_{1}d_2)\right].$$
To further higher order of $\epsilon$, substituting Eqs.
(\ref{7}), (\ref{8}) and (\ref{30})$-$(\ref{33}) into Eqs.
(\ref{2})$-$(\ref{4}), and then taking the terms containing
$\epsilon^{2}$ from  Eq. (\ref{2}) and  the z-component of Eq.
(\ref{3}), and $\epsilon^{3/2}$ from Eq. (\ref{4}), we obtain
another set of equations:
\begin{eqnarray}
&&\hspace*{-8mm}\frac{\partial n_h^{(1)}}{\partial
\tau}-V_p\frac{\partial n_h^{(3)}}{\partial\xi}+l_x\frac{\partial
u_{hx}^{(2)}}{\partial\xi}+l_x\frac{\partial}{\partial\xi}(n_h^{(1)}u_{hx}^{(1)})\nonumber\\
&&\hspace*{-8mm}+l_y\frac{\partial
u_{hy}^{(2)}}{\partial\xi}+l_y\frac{\partial}{\partial\xi}(n_h^{(1)}u_{hy}^{(1)})+l_z\frac{\partial
u_{hz}^{(3)}}{\partial\xi}\nonumber\\
&&\hspace*{-8mm}+l_z\frac{\partial}{\partial\xi}(n_h^{(1)}u_{hz}^{(2)})+l_z\frac{\partial}{\partial\xi}(n_h^{(2)}u_{hz}^{(1)})=0,
\label{37}\\
&&\hspace*{-8mm}\frac{\partial u_{hz}^{(1)}}{\partial
\tau}-V_p\frac{\partial
u_{hz}^{(3)}}{\partial\xi}+l_z\frac{\partial}{\partial\xi}(u_{hz}^{(1)}u_{hz}^{(2)})+l_z\frac{\partial\phi^{(3)}}{\partial\xi}=0,
\label{38}\\
&&\hspace*{-8mm}\frac{\partial^2\phi^{(1)}}{\partial\xi^2}=(\mu_{0}c_1-\mu_{1}d_1)\phi_3+2(\mu_{0}c_2 \nonumber\\
&&\hspace*{-8mm}-\mu_{1}d_2)\phi^{(1)} \phi^{(2)} -n_h^{(3)}.
\label{39}
\end{eqnarray}
Now, combining Eqs. (\ref{37})$-$(\ref{39}), we finally obtain the
mK-dV equation:
\begin{eqnarray}
&&\hspace*{-10mm}\frac{\partial\phi^{(1)}}{\partial \tau} +
\alpha_1 \alpha_3\phi^{(1)2} \frac{\partial \phi^{(1)}}{\partial
\xi}+\alpha_2\alpha_3\frac{\partial^3 \phi^{(1)}}{\partial
\xi^3}=0. \label{DIAmK-dV}
\end{eqnarray}
where
\begin{eqnarray}
&&\alpha_1=\frac{15l_z^6}{2V_p^6},
\label{40}\\
&&\alpha_2=1+\frac{(1-L_z^2)}{\alpha^2},
\label{41}\\
&&\alpha_3=\frac{V_p^3}{2L_z^2}. \label{42}
\end{eqnarray}
To solve this mK-dV equation, We consider a frame
$\xi=\eta-u_{0}T$ (moving with speed $u_{0}$). The stationary
solitary wave solution of the MK-dV equation [Eq.
(\ref{DIAmK-dV})] is given by
\begin{eqnarray}
{\rm \phi^{(1)}}=\rm \phi_m{\rm[sech}(\frac{\xi}{\varpi})],
\label{solMK-dV}
\end{eqnarray}
where  $\phi_m=\sqrt{6u_{0}/\alpha_1 \alpha_3}$ is the amplitude,
and $\varpi={\sqrt{u_{0}/\alpha_3}}$ is the width of HIA SWs.
\begin{figure}[t!]
\centerline{\includegraphics[width=8cm]{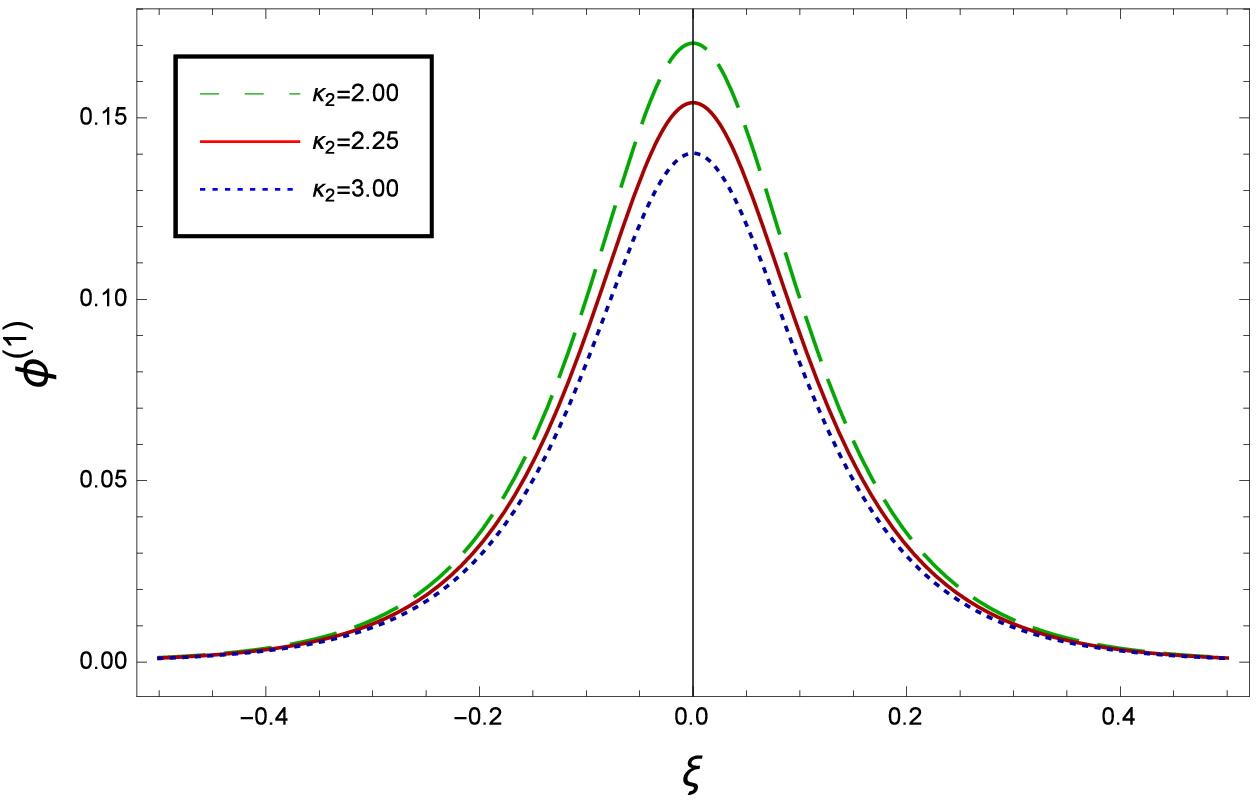}} \caption{(Color
online) The ESPPs with $\phi^{(1)}>0$ for $u_0=0.01$, $\sigma =
0.25$, $\mu=2.98$, $\kappa_1=20$, $\delta=15$, $\alpha =0.5$,
$\kappa_2=2.00$ (dashed curve), $\kappa_2=2.25$ (solid curve),
and $\kappa_2=3.00$ (dotted curve).} \label{Fig8}
\end{figure}
\begin{figure}[t!]
\centerline{\includegraphics[width=8cm]{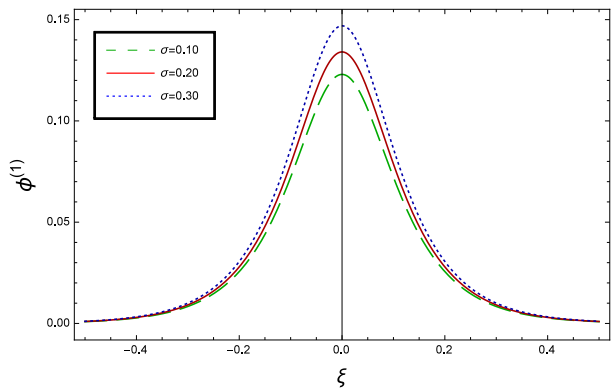}} \caption{(Color
online) The ESPPs with $\phi^{(1)}>0$ for $u_0=0.01$, $\delta =
15$, $\mu=2.98$, $\kappa_1=20$, $\kappa_{2}=3$, $\alpha =0.5$,
$\sigma=0.10$ (dashed curve), $\sigma=0.20$ (solid curve), and
$\sigma=0.30$ (dotted curve).} \label{Fig9}
\end{figure}
\begin{figure}[t!]
\centerline{\includegraphics[width=8cm]{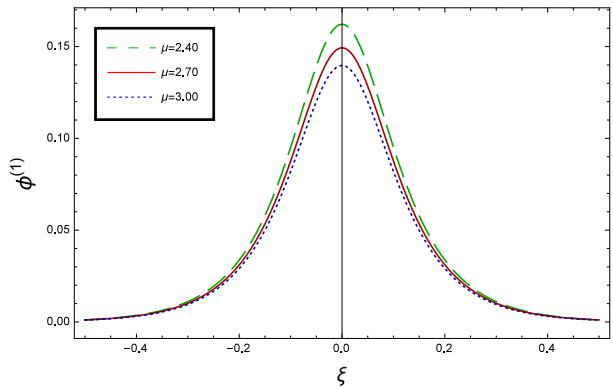}} \caption{(Color
online) The ESPPs with $\phi^{(1)}>0$ for $u_0=0.01$, $\delta =
15$, $\sigma=0.25$, $\kappa_1=20$, $\kappa_{2}=3$, $\alpha =0.5$,
$\mu=2.40$ (dashed curve), $\mu=2.70$ (solid curve), and
$\mu=3.00$ (dotted curve).} \label{Fig10}
\end{figure}

To identify the salient features (viz. polarity, amplitude, and
width) of the ESPPs, we have numerically analyzed the solution of
the mK-dV equation [Eq. (\ref{solMK-dV})] for different plasma
parametric regimes. The results are depicted in Figs.
\ref{Fig8}$-$\ref{Fig10}, which clearly indicates that (i)the
mK-dV equation admits solitary wave solution with $\phi^{(1)}>0$
only; (ii) the amplitude and width of the ESPPs decrease with the
increase in ($\kappa_2$) as shown in Fig. \ref{Fig8}; (iii) the
amplitude and width  of the ESPPs increase with the increase in
$\sigma$ as shown in Fig. \ref{Fig9}; (iv) the amplitude and width
of the ESPPs decrease with the increase in $\mu$ as shown in Fig.
\ref{Fig10}.
\section{Discussion}
\label{Sec:4} We have considered a magnetized plasma system
consisting of inertial heavy ions and  kappa distributed hot
electrons and hot positrons of two distinct  temperatures. We
have derived  the magnetized K-dV and mK-dV-type partial
differential equations by using the reductive perturbation method
to investigate  the basic features (i.e.polarity, amplitude, and
width) of such a plasma system. The magnetized K-dV  and MK-dV
equations are solved to set out the fascinating features of HIA
SWs. Then these solutions are analyzed by taking the effect of
different plasma parameters. The results, which have been
obtained from this theoretical investigation, can be pin-pointed
as follows:
\begin{enumerate}
\item{The K-dV equation admits HIA SW solutions with either $\phi^{(1)}>0$ (compressive) or
$\phi^{(1)}<0$ (rarefactive). The polarity of the HIA SWs depends
on the critical value $\mu_c$ (where $\mu_c=2.98$ for
$\kappa_1=20$, $\kappa_2=3$, $\delta=15$, $\sigma=0.25$, and
$\alpha=0.5$). On the other-hand, the MK-dV equation admits only
HIA SW solution with $\phi^{(1)}>0$ (compressive).}
\item{The K-dV equation is no longer valid at
$A_1\simeq 0$ because the amplitude of the K-dV solitons become
infinitely large (for $A_1=0$), which has been avoided by deriving
MK-dV equation to study more highly nonlinear HIA SWs.}
\item{The amplitude and width of both positive and negative HIA SWs(obtained from the numerical analysis of the solution of the K-dV equation)
increase with the increase in $T_e$ and $n_{p0}$ but decrease
with the increase in $T_p$, $n_{eo}$, and $\kappa_2$.}
\item{The width of the K-dV solitons decreases with the increase in $\alpha$, and increases
(decreases) with the increase in $\delta$ for its lower (upper)
range.}
\item{The amplitude and the width of the MK-dV HIA SWs decrease with the increase in $\kappa_2$, $T_p$, and $n_{e0}$, but increase with the increase in $T_e$ and $n_{p0}$.}
\end{enumerate}
Therefore, we hope that our present investigation would contribute
to understand the prime features (i.e. polarity, amplitude, and
width) of the electrostatic disturbances of HIA SWs in a
magnetized plasma system. Our findings would be useful to study
nonlinear structures in space (viz. peculiar velocities of galaxy
clusters, cluster explosions, active galactic nuclei, pulsar
magnetosphere, ionosphere \cite{Bremer1996}, Saturn's
magnetosphere \cite{Baluku2012}, solar wind \cite{Pierrard1996},
etc.) as well as laboratory plasma conditions
\cite{Tribeche2009,Surko1989,Abdullah1995,Kurz1998,Greaves2002}
(viz. semiconductor plasmas \cite{Shukla1986}) containing heavy
ions where the effect of two temperature superthermal electrons
and positrons play a crucial role.

\end{document}